**ColourQuant: a high-throughput technique to extract and quantify colour phenotypes from plant images**

Mao Li[1*], Margaret H. Frank[2*], and Zoë Migicovsky[3]

[1]Donald Danforth Plant Science Center, 975 North Warson Road, St. Louis, MO 63132-2918, USA
[2]Plant Biology Section, School of Integrative Plant Science, Cornell University, Ithaca, NY 14853
[3]Department of Plant and Animal Sciences, Faculty of Agriculture, Dalhousie University, Truro, Nova Scotia, B2N 5E3, Canada

*Co-corresponding authors: Mao Li mli@danforthcenter.org; Margaret Frank mhf47@cornell.edu

**Running Head: ColourQuant: high-throughput colour phenotyping**

**Abstract**

Colour patterning contributes to important plant traits that influence ecological interactions, horticultural breeding, and agricultural performance. High-throughput phenotyping of colour is valuable for understanding plant biology and selecting for traits related to colour during plant breeding. Here we present ColourQuant, an automated high-throughput pipeline that allows users to extract colour phenotypes from images. This pipeline includes methods for colour phenotyping using mean pixel values, Gaussian density estimator of Lab colour, and the analysis of shape-independent colour patterning by circular deformation.

**Keywords:**

Colour phenotyping; high-throughput image acquisition; colour patterning; continuous colour distribution; shape-independent colour quantification

**1. Introduction**

Colour patterning contributes to important ecological, horticultural, and agricultural traits. Developing high-throughput (HT) phenotyping methods for colour analysis is essential for furthering our understanding of plant biology and providing accurate, quantitative information for plant breeding.

Morphological diversity in flower colour plays a significant role in determining pollinator recruitment, and as a result, pollinator preference may result in selection for flower colours. For example, across 206 Australian angiosperm species, flowers pollinated by birds differed significantly in colour from those visited by insects *(1)*. Pollinators may also exert selective pressure on colour variation among close relatives, as evidenced by the hummingbird pollinated clade Iochrominae (Solanaceae) *(2)*.

In addition to driving plant-pollinator relationships, colour is an essential component of the ornamental plant industry and has a direct influence on the commercial value of given cultivars *(3)*. Indeed, the desire for new colours has been a major driver behind the use of biotechnology in ornamental horticulture, especially for the cut flower industry *(4)*.

Food colour can significantly influence flavour perception *(5, 6)*, steering consumer choices. For fresh market produce, consumers prefer bright colours, which can indicate freshness and desirable nutrient content *(7)*. Even after the initial purchasing decision, colour may influence consumer perception of fruit flavour *(8)*. However, traditional assessment of seedlings for colour may be time-consuming for large numbers of plants and vary based on observer. Thus, efficient and accurate phenotyping of colour and colour patterning may serve as an important tool for plant breeding.

Quantitative colour measurements may be used directly for culling plants without a desirable trait or, in instances where the colour is not apparent at the seedling stage (e.g. fruit on trees), the data can be used instead for genetic mapping. Techniques such as linkage mapping and genome-wide association studies (GWAS) connect phenotype data with genotype data to uncover genetic markers correlated with a trait of interest. Genetic mapping is improved by precise, quantitative data *(9)*. Early screening of plants using genetic markers allows the breeder to reduce the number of plants that are propagated without a trait of interest and therefore is especially cost-effective in perennial crops that have a lengthy juvenile phase, such as apples and grapes *(10)*. As a result, genomics-assisted breeding for colour traits including peach blush *(11)* and sweet cherry fruit colour *(12)* are already underway.

HT phenotyping can also be used to efficiently detect and diagnose pathogen spread in diseased plants *(13)*, facilitating genetic mapping of disease resistance. Genomics-assisted breeding of disease resistance eliminates the need for the time-consuming and expensive task of inoculating plants. Digital imaging not only improves ease of scoring for infection, but also allows for a quantitative measurements of characteristics that would otherwise be missed. For example, a recent study of *Arabidopsis thaliana* infected with *Botrytis cinerea* re-analyzed images from a previous GWAS for visual traits, including colour, finding that some resistance genes impacted colour, but not the shape or size of lesions *(14)*.

Among the benefits of HT colour phenotyping is its potential to dramatically improve our characterization and understanding of plant diversity, such as plant-pollinator relationships, and have a direct impact on plant breeding for important traits including appearance and disease. Here we present ColourQuant, methods for automated HT colour phenotyping using mean pixel values, Gaussian density estimation of Lab colour, and the analysis of shape-independent colour patterning by circular deformation.

**2. Materials**

1. Flatbed scanner (example, Epson Perfection V550 Scanner) for flat images, camera and lightbox for three-dimensional objects
2. Colour card (example, Kodak KOCSGS color separation guide)
3. Matlab (https://github.com/maoli0923/ColourQuant)

**3. Methods**

3.1. Image Acquisition

1. For flat objects such as leaves, place samples on a flatbed scanner with a colour card in the corner.
2. For three-dimensional objects, such as fruit, images may be acquired by placing the samples inside of a lightbox with a colour card in the corner, and photographing with a camera.
3. A few rounds of sample images should be collected in order to optimize lighting and resolution. It is useful to have enough light to capture details while reducing glare. Minimizing shadows around the edges of your samples will also help with streamlining the downstream image processing steps.
4. Colour images can be saved in a variety of file formats. Lossless compression methods retain complete pixel information, these include: TIF LZW and PNG file formats. JPG files are produced using Lossy compression, which reduces pixel information, making the files smaller, but less informative. Lossless compression is generally preferred for colour image analysis; however, Lossy compression works in most cases and takes up less computer storage. The right file format will depend on the size of the experiment, hard drive space, and desired experimental output.
5. Colour correct images. One method is to perform white balance for the image. In the example pictured, first extract and average R, G, and B values (denoting as avgR, avgG, and avgB) for the white swatch in the Kodak KOCSGS color separation guide. Then adding (255-avgR), (255-avgG), and (255-avgB) to R, G, and B of all the pixels in the image. As a result, the RGB value of the white swatch on the colour guide is equal to 255 and the image colour is white balanced. (**Fig. 1, code lines 5-29**)

3.2. Object Segmentation

To separate the object (e.g leaf, fruit) from the background, first convert the image into a binary image in which object and background are in white and black (or black and white), respectively. Many different segmentation methods exist, such as adaptive thresholding or learning algorithms *(15)*. In addition, parameters may need to be adapted based on factors such as brightness and contrast of the image with respect to the background, surrounding shadows along the sample margins, and image quality. One object segmentation method that generally works well for scanned as well as photographed objects is outlined below (**code lines 31-69**), and displayed in **Fig. 1**.

1. Extract the RGB matrix from the image and convert it into hue-saturation-value (HSV) format.
2. In HSV, most background pixels become grey and it is possible to set a threshold that separates grey values from true object values. In our example, this value was Saturation>0.15, but it will need to be customized for each experiment (**code line 39**).
3. If there are multiple objects in an image (e.g coleus image in **Fig. 1**), detect each connected component and segment out the object of interest (e.g. large enough objects, **code lines 44-51**) for color analysis.

3.3. Color analysis

The first step in colour analysis described here is to convert the color matrices from RGB to CIE Lab color (**code lines 78-85**). CIE Lab color is a continuous color space that consists of three descriptors: L = "lightness", a = "green to magenta", and b = "blue to yellow," displayed in Fig. 1.

3.3.1. Mean and variance

For objects with nearly solid colours or relatively simple patterns, calculating the mean and variance for L, a, b color values for each image is informative (**code lines 72-91**). These data can be summarized visually using a scatter plot with each object (e.g. apple fruit) displayed (**Fig 2**).

3.3.2. Gaussian density estimator

For objects with complex color patterns, a robust and more comprehensive measurement tool, such as a Gaussian density estimator (GDE), needs to be applied.

1. Treat 3D Lab colour matrices as 3D point clouds with coordinates (L, a, b).
2. To reduce the amount of time needed for the computation, we find the extreme values for L, a, and b through 3D point clouds of the population (**code lines 88-90, 95**). Then working 3D space can be bounded by a box with ranges based on the extreme values L, a, and b. In our example, L ranges between -10 and 110, a ranges between -40 and 50, and b ranges between -30 and 74 (**code lines 94-97**).
3. Extract colour distribution and frequency for each image by applying a GDE to the Lab point cloud (**code lines 98-112**). The GDE directly estimates density from the point cloud data, thus, it is a function defined on a 3D space (depicted in **Fig. 1**).
4. The GDE descriptor captures statistical colour distributions; however, it does not provide information regarding spatial patterning. To capture spatial color information, the object can be segmented into distinct zones. One method is to define these zones based on normalized pixel distances (**code lines 114-146**). For example, the "border", defined as the outer 15% of pixels from the leaf boundary to the centroid, the "center", defined as the inner 75% of pixels from the centroid to the boundary, and "full", defined as the entire colour matrix. These zones should be customized for each study (**code lines 134-135**). The distance between any two objects is calculated by

$$D = \sqrt{d^2_{full} + d^2_{border} + d^2_{center}}$$

This calculation determines the difference in colour patterns between two objects based on their similarity across all zones (**code lines 152-156**)

### 3.3.3. Circular deformation

To examine the impact of colour while reducing the effect of shape, it is possible to deform each object (e.g. coleus leaf) into a disk using thin-plate spline (TPS) interpolation *(16)*; algorithm is from *(17)*.

1. Align all the objects to the same orientation (e.g. rotating the leaf so that the tip is on the top and base is on the bottom). This could be achieved by aligning a few manually or automatically labeled landmarks (e.g. leaf tip and base, **code lines 163-176**).
2. Normalize the object so that the square root of the average squared distances of all the points on the outline to the center is 1 (**code line 177-178**).
3. Set the points on the outline as control points
4. Use TPS to deform the object to force the control points to be the points on the circle with radius 1 (**Fig. 3A**) and save the image with a transparent background (**code lines 182-199**).
5. Resize the circular image with a fixed dimension (e.g. 70x70). Extract Lab colors form a 14700 dimensional vector (4900 pixels, each has 3 values, **code lines 202-212**).
6. It is possible to perform principal component analysis on these vectors (**Fig. 3B**) and get eigen colors (**Fig. 3C**) that show color pattern variation (**code lines 214-232**).

## 4. Conclusion

This method enables the efficient extraction of quantitative color distribution and patterning from a large set of samples. It can be applied to virtually any subject, and flexibly adapted to study different color patterns and investigate color patterning irrespective of sample shape. The output from this method includes a table of color values that correspond to each sample or sub-sample, for which there are numerous visualization and statistical packages built in R and Matlab that can be used to analyze and plot the data.

**Acknowledgements**

Z.M. was supported by NSF 1546869. M.H.F. is supported through startup funds from Cornell University's College of Agriculture and Life Science. The coleus samples used in Figs 1 and 3 were imaged from Dr. David Clark's breeding program at the University of Florida Gainesville.

**Figure captions**

Figure 1. Overview of ColourQuant pipeline. Flat samples can be imaged using a scanner or digital camera, whereas three-dimensional samples are easiest to capture on a copy stand (A). For both approaches, the inclusion of a colour card is essential for performing post-acquisition colour balance. Colour thresholding can be used to segment samples from the surrounding background (B), and then colour values for isolated samples can be extracted and quantified using a three step process. First, pixels are converted from RGB to Lab continuous colour space, next this spaces is plotted in a three-dimensional point cloud, and finally a Gaussian Density Estimator function is applied to the point cloud, in order to quantify the colour composition of the sample (C).

Figure 2. Mean "L" and "a" pixel values displayed as a Scatter plot of apple samples. In continuous colour space, L represents the light-to-dark spectrum of values and a represents magenta to green space. In this example, apple samples spread from light green to deep red across the Scatter plot.

Figure 3. Example of the thin plate spline method applied to coleus leaves to analyze colour composition independent of sample shape. In this example coleus leaves are deformed into circles using the thin plate spline method (A), and then plotted into PCA space based on their global pixel composition (B). Eigen

leaves that explain the largest contribution to colour variance across the sample population can be extracted from this analysis (C).

Figure 1: Overview of ColourQuant pipeline

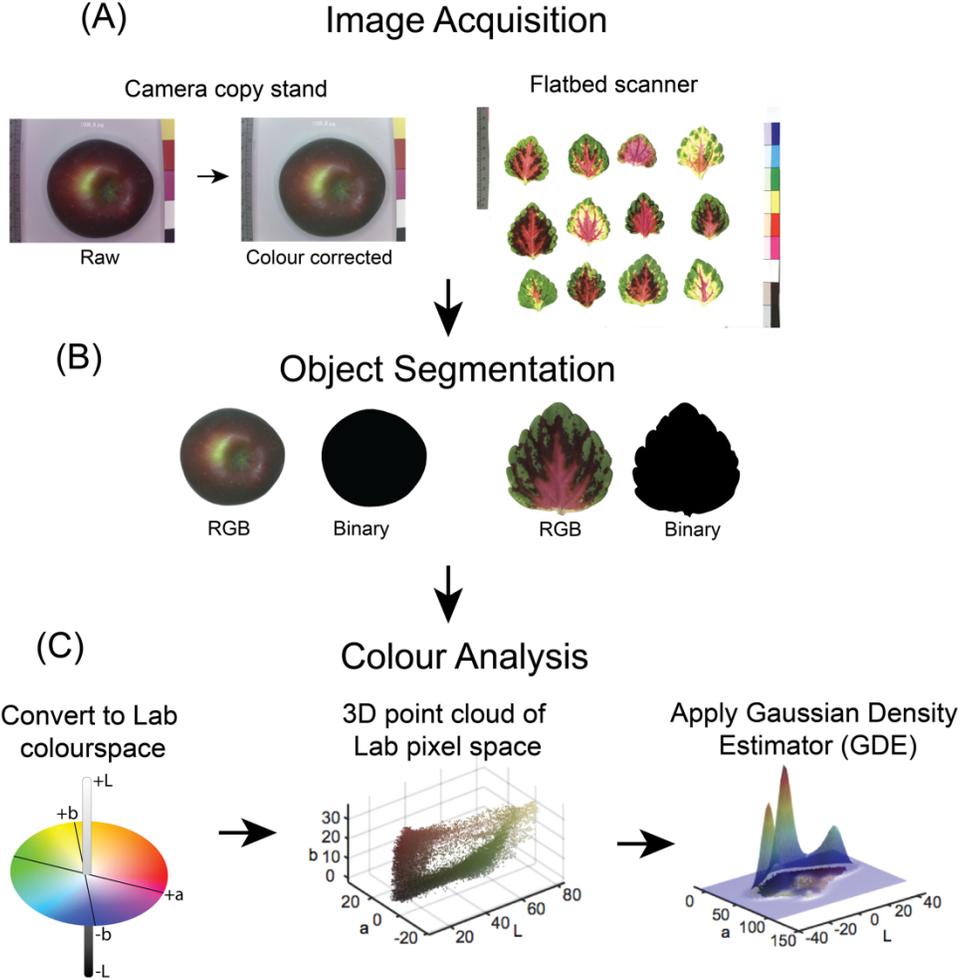

Figure 2: Mean "L" and "a" pixel values displayed as a Scatter plot of apple samples

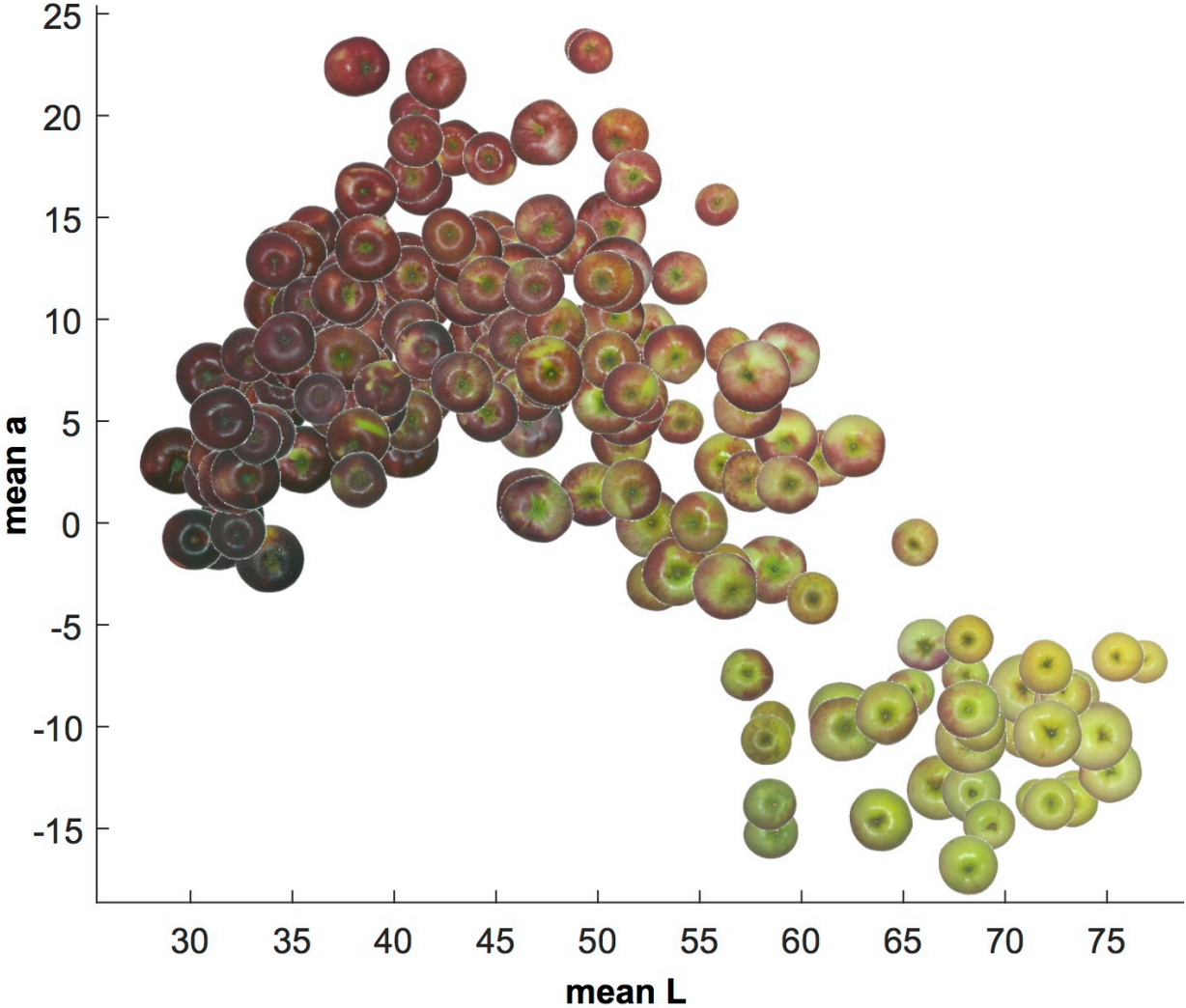

Figure 3: Example of the thin plate spline method applied to coleus leaves to analyze colour composition independent of sample shape

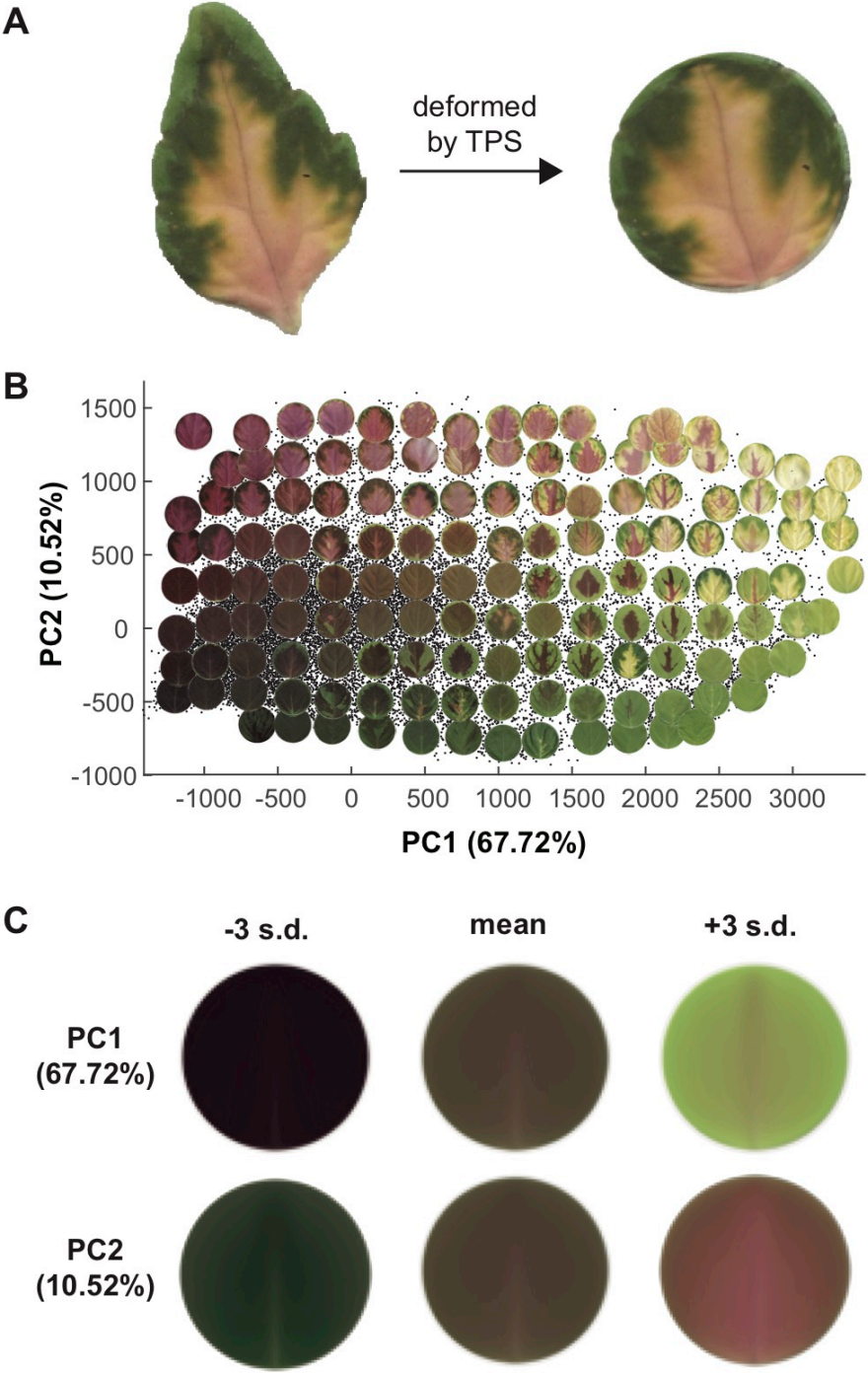